\documentclass[journal]{IEEEtran}

\usepackage[T1]{fontenc}
\usepackage[utf8]{inputenc}
\usepackage{amsmath, amssymb, bm}
\usepackage{booktabs}           
\usepackage{array}              
\usepackage{xcolor}
\usepackage{titlesec}           
\usepackage{enumitem}           
\usepackage{mdframed}           
\usepackage{multirow}
\usepackage{etoolbox}
\usepackage{glossaries}
\usepackage{glossaries-extra}
\usepackage{amsmath,amssymb,amsfonts}
\usepackage{algorithm}
\usepackage{algorithmicx}
\usepackage{graphicx}
\usepackage{textcomp}
\usepackage{array}
\usepackage{footnote} 
\usepackage{tablefootnote}      
\usepackage{xcolor}
\usepackage{adjustbox}
\usepackage{booktabs} 
\usepackage{bm}
\usepackage{algpseudocode}
\usepackage{tabularx}
\usepackage{bm}
\usepackage[dvipsnames]{xcolor}
\usepackage[table]{xcolor}
\usepackage{soul}  
\usepackage[caption=false,font=normalsize,labelfont=sf,textfont=sf]{subfig}
\usepackage{mathtools}
\usepackage{booktabs}
\usepackage{makecell} 
\usepackage[colorlinks=true, allcolors=blue]{hyperref}
\usepackage{cleveref}
\usepackage{tikz}

\def\BibTeX{{\rm B\kern-.05em{\sc i\kern-.025em b}\kern-.08em
    T\kern-.1667em\lower.7ex\hbox{E}\kern-.125emX}}
    
\hypersetup{
    colorlinks=true,
    linkcolor=black,     
    citecolor=blue,      
    urlcolor=blue,
    pdftitle={Unified GNN Solver Paper}, 
}
\usetikzlibrary{positioning}
\usetikzlibrary{shapes.geometric, arrows.meta, positioning, calc}
\newcommand{\abs}[1]{\left|#1\right|}

\definecolor{col_res}{RGB}{200,100,20}
\definecolor{col_gin}{RGB}{35,100,180}
\definecolor{col_sag}{RGB}{25,130,65}
\definecolor{col_hdr}{RGB}{40,40,55}
\definecolor{col_box}{RGB}{245,246,250}
\definecolor{col_rule}{RGB}{180,180,195}

\hypersetup{
  colorlinks=true,
  linkcolor=col_hdr,
  urlcolor=col_gin,
  citecolor=col_gin
}

\newmdenv[
  backgroundcolor=col\_box,
  linecolor=col\_rule,
  linewidth=0.8pt,
  roundcorner=4pt,
  innerleftmargin=12pt,
  innerrightmargin=12pt,
  innertopmargin=10pt,
  innerbottommargin=10pt,
  skipabove=8pt,
  skipbelow=8pt
]{eqframe}

\setabbreviationstyle[acronym]{long-short}

\newacronym{ip}{IP}{Interior Point}
\newacronym{nr}{NR}{Newton--Raphson}
\newacronym{wls}{WLS}{Weighted Least Squares}

\newacronym{pf}{PF}{Power Flow}
\newacronym{opf}{OPF}{Optimal Power Flow}
\newacronym{se}{SE}{State Estimation}

\newacronym{gnn}{GNN}{Graph Neural Network}
\newacronym{gat}{GAT}{Graph Attention Network}
\newacronym{nn}{NN}{Neural Network}
\newacronym{cnn}{CNN}{Convolutional Neural Network}
\newacronym{mpnn}{MPNN}{Message Passing Neural Network}
\newacronym{ae}{AE}{Autoencoder}
\newacronym{vae}{VAE}{Variational Autoencoder}
\newacronym{vgae}{VGAE}{Variational Graph Autoencoder}
\newacronym{gru}{GRU}{Gated Recurrent Unit}
\newacronym{gine}{GINE}{Graph Isomorphism Network with Edge features}
\newacronym{gin}{GIN}{Graph Isomorphism Network}
\newacronym{mlp}{MLP}{Multilayer Perceptron}
\newacronym{gcn}{GCN}{Graph Convolutional Neural Networks}
\newacronym{resgated gcn}{ResGated GCN}{Residual Gated Graph Neural Network}

\newacronym{fm}{FM}{Foundational Model}
\newacronym{kcl}{KCL}{Kirchhoff Current Law}
\newacronym{mse}{MSE}{Mean Squared Error}
\newacronym{nrmse}{NRMSE}{Normalized Root Mean Square error}

\newacronym{hgnn}{HGNN}{Heterogeneous GNN}

\begin{document}

\title{Unified Heterogeneous Graph Neural Network solver for Power Flow, Optimal Power Flow and State Estimation}

\author{
Ferran~Bohigas-Daranas,
Hamid~Latif-Martínez, ~Eduardo Prieto-Araujo,~\IEEEmembership{Member~IEEE}, ~Oriol~Gomis-Bellmunt,~\IEEEmembership{Member~IEEE}, ~Pere Barlet-Ros,~\IEEEmembership{Member~IEEE}
\thanks{The research has been funded by projects Daedalos(Horizon Europe research grant agreement No 101172829), GRAPHS4SEC (grant PCI2023-145974-2 funded by MICIU/AEI/10.13039/501100011033) and BLOSSOMS (grant PID2024-158530OB-I00, by MI-CIU/AEI/10.13039/501100011033/ and ERDF/EU). The work of P.Barlet, O.Gomis-Bellmunt and E.Prieto-Araujo was supported by the Agència de Gestió d’Ajuts Universitaris i de Recerca (AGAUR) through the ICREA Acadèmia programme, and by the Departament de Recerca i Universitats of the Generalitat de Catalunya. E. Prieto-Araujo is a member of the Serra Húnter Programme.}

\vspace{-1.5em}
}

\markboth{Journal of transactions on power systems}%
{Shell \MakeLowercase{\textit{et al.}}: Bare Demo of IEEEtran.cls for IEEE Journals}

\maketitle

\begin{abstract}
Power Flow (PF), Optimal Power Flow (OPF), and State Estimation (SE) are fundamental problems in power system analysis, but solving them is computationally expensive. Graph Neural Networks (GNNs) have been proposed as fast surrogates, yet existing solvers are trained for a single problem at a time, producing narrow models that must be rebuilt for each new task.

We propose a more general approach: a single Heterogeneous Residual Gated Graph Convolutional Network that solves all three problems with one shared backbone. Rather than learning one mapping, the model learns a reusable representation of how the network behaves, from which PF, OPF, and SE can each be estimated. Trained jointly on the three problems across diverse topologies and loading conditions, and evaluated on the IEEE 14-bus and 118-bus systems, the shared model matches the accuracy of task-specific GNN solvers and stays robust on unseen loading levels and topologies.

These results show that a single model can capture the basic operation of a power network and serve several analysis tasks at once — a first step toward a foundation model for power systems.
\end{abstract}
\glsresetall


\IEEEpeerreviewmaketitle

\glsresetall

\section{Introduction}
\label{sec:introduction}
 
\IEEEPARstart{T}{he} rapid transformation of modern power grids, driven by the integration of renewable energy sources and the deployment of advanced metering infrastructure, has significantly increased the complexity of grid monitoring and control. Fundamental power system tasks~\cite{grainger_power_1994}, namely \gls{pf}, \gls{opf}, and \gls{se}, remain the cornerstones of grid operations. While traditional iterative methods such as \gls{nr}~\cite{tinney_power_1967}, \gls{ip}~\cite{capitanescu_critical_2016}, and \gls{wls}~\cite{schweppe_power_1970} are mathematically robust, their computational burden scales super-linearly with system size \cite{crow2015computational}. Furthermore, the increasing variability of distributed energy resources (DERs) demands real-time or near-real-time solutions that traditional solvers struggle to provide under strict latency constraints.
 
In recent years, Deep Learning (DL) has emerged as a promising alternative for approximating these complex power system mappings. Early applications utilized Multilayer Perceptron (MLP) and Convolutional Neural Networks (CNN) ~\cite{marot_learning_2019,pham_neural_2022}. However, these architectures are fundamentally limited when applied to power systems: \glspl{mlp} fail to exploit the topological structure of the grid, while CNNs are restricted to regular, Euclidean data structures and cannot naturally represent the irregular connectivity of transmission networks.
 
Given that power systems are inherently non-Euclidean graphs, defined by buses (nodes) and lines/transformers (edges), \glspl{gnn} \cite{scarselli_graph_2009, gilmer2017neuralmessagepassingquantum,battaglia2018relationalinductivebiasesdeep} have emerged as a natural candidate for power system applications. The message-passing mechanism of GNNs directly mirrors the way physical laws such as \gls{kcl} propagate information across the network, introducing a topological inductive bias that aligns the learning process with the relational structure of the electrical grid. GNN-based approaches have since demonstrated strong results for individual power system tasks: \gls{pf} solvers based on GNNs
\cite{lopez-garcia_power_2023, lin_powerflownet_2024} have achieved near-\gls{nr} accuracy at a fraction of the computational cost; GNN-based \gls{opf} methods \cite{owerko_optimal_2020,donon_neural_2020, lopez-garcia_optimal_2024,varbella_physics-informed_2024} have shown the ability to respect operational constraints while accelerating dispatch decisions; and GNN architectures for \gls{se}~\cite{saur_graph_2024,kundacina_state_2022} have demonstrated robustness to measurement noise and partial observability. Despite these advances, a key limitation persists: early and current GNN variants often lack the discriminative power required to handle the high-precision requirements of power system physics~\cite{xu_how_2019}.
This work addresses this limitation by utilizing the \gls{resgated gcn} \cite{bresson_residual_2018}, as the basic computational cell of the proposed architecture (Fig.\ref{fig:gnn_architecture}), providing stronger graph-distinguishing capability than prior Graph Convolutional Networks (GCN) and Graph Attention Networks (GAT) based approaches.
 
Despite the progress in GNN-based power system solvers, existing literature consistently treats \gls{pf}, \gls{opf}, and \gls{se} as isolated problems, with separate models trained and deployed for each task. This fragmentation ignores the shared underlying physics and topological constraints common to all three problems, results in redundant training pipelines, and prevents the development of a unified, grid-aware representation. The recent emergence of foundational model initiatives for power systems, notably the open-source initiatives GridFM~\cite{gridfm_gridfm_2025} and AI.grids~\cite{cresym_aigrids_nodate}, and more recently by industry research groups~\cite{weiwei_yang_gridsfm_2026} confirms that the community recognizes the need to move beyond task-specific architectures. However, to date no peer-reviewed architecture has demonstrated simultaneous solution of \gls{pf}, \gls{opf}, and \gls{se} within a single trained model on standard benchmark systems. This paper directly addresses that gap.

The primary contributions of this work are:
 
\begin{itemize}[leftmargin=*]
    \item \textit{Unified multi-task architecture:} A unified heterogeneous GNN architecture trained to solve \gls{pf}, \gls{opf}, and \gls{se} simultaneously, including topological perturbations, so that the model learns a generalized representation of power system physics rather than a task-specific input-output mapping, achieving accuracy comparable to task-specific GNNs while eliminating the need for separate per-task training pipelines. To the best of the authors' knowledge, this is the first peer-reviewed architecture to demonstrate this on standard IEEE benchmark systems.
    \item \textit{Multi-problem loss:} A composite loss function for the three problems (\gls{pf}, \gls{opf}, \gls{se}) simultaneously, with a problem-conditioned masking strategy that prevents cross-problem gradient interference.
    \item \textit{Architectural hyperparameter tuning and ablation study:} A systematic hyperparameter tuning and ablation that quantifies the individual contribution of each element of the propose architecture, providing evidence-based architectural guidance for future unified power system models.
\end{itemize}
 
The remainder of this paper is organized as follows. Section~\ref{sec:electrical background} introduces the mathematical formulation of \gls{pf}, \gls{opf} and \gls{se}. Section~\ref{sec:gnn background} reviews the GNN foundations underpinning the proposed architecture. Section~\ref{sec:proposed methodology} describes the proposed architecture, training objective, and data generation pipeline. Section~\ref{sec:results} presents the experimental setup, numerical results, and discussion on the IEEE 14-bus and 118-bus test systems \cite{ieee118bus}, including an analysis of scalability and the path towards foundational power system models. Section~\ref{sec:conclusion} concludes the paper.

\section{Electrical Background}
\label{sec:electrical background}

The power system is modeled as a network of buses interconnected
by transmission lines and transformers. Each bus is characterized
by four electrical quantities: voltage magnitude $\abs{V}$,
voltage phase angle $\delta$, net active power injection $P$, and
net reactive power injection $Q$, where \emph{net} denotes the
algebraic difference between local generation and consumption.

Buses are classified into three types according to which variables  are specified as inputs and which are computed as outputs. The slack bus (or reference bus) is typically the terminal of the largest generating unit; its voltage magnitude and angle are held fixed, and it absorbs any residual active and reactive power mismatch in the system. PV buses correspond to generator terminals at which active power output and voltage magnitude are prescribed, while reactive power output and voltage angle are unknowns. PQ buses represent load nodes, where both active and reactive power demands are specified and the voltage phasor is to be determined.

For each bus $i$ in an $n$-bus system, the complex power injection is related to the nodal voltage phasor and the injected current by:
\begin{equation}
  S_i = P_i + jQ_i = V_i  I_i^{*},
  \label{eq:complex_power}
\end{equation}
where $S_i$ is the bus power, $P_i$ is the bus active power, $Q_i$ is the bus reactive power, $V_i$ is the bus voltage and $I_i^*$ is the complex conjugate of the current injection at bus $i$. 

Applying Ohm's law through the nodal admittance matrix ${Y}_{\text{bus}}$, whose 
(i,j)-th element is expressed in polar form as $Y_{ij} = |Y_{ij}|e^{j\theta_{ij}}$
, with $\theta_{ij} = \arg(Y_{ij}$):
\begin{equation}
  I_i = \sum_{j=1}^{n} Y_{ij}\,V_j.
  \label{eq:ohm}
\end{equation}
Substituting~\eqref{eq:ohm} into~\eqref{eq:complex_power} and separating real and imaginary parts yields the fundamental power flow equations:
\begin{align}
  P_i &= \sum_{j=1}^{n} \abs{V_i}\abs{V_j}\abs{Y_{ij}}
          \cos\!\left(\theta_{ij} - \delta_i + \delta_j\right),
  \label{eq:pflow} \\
  Q_i &= -\sum_{j=1}^{n} \abs{V_i}\abs{V_j}\abs{Y_{ij}}
          \sin\!\left(\theta_{ij} - \delta_i + \delta_j\right),
  \label{eq:qflow}
\end{align}
where $Y_{ij}$ is the $(i,j)$-th element of the admittance matrix and $\delta_i$, $\delta_j$ are the voltage angles at buses $i$ and $j$, respectively. 

Nodal power balance, derived from \gls{kcl}, at each bus  $i \in \mathcal{N}$:
\begin{align}
  P_{Gi} - P_{Di} - P_i(V,\delta) &= 0, \quad \forall i \in \mathcal{N},
  \label{eq:pbalance} \\
  Q_{Gi} - Q_{Di} - Q_i(V,\delta) &= 0, \quad \forall i \in \mathcal{N},
  \label{eq:qbalance}
\end{align}
where $P_{Gi}$, $Q_{Gi}$ are active and reactive generation, $P_{Di}$, $Q_{Di}$ are the corresponding demands, and $P_i(V,\delta)$, $Q_i(V,\delta)$ are computed from~\eqref{eq:pflow}-\eqref{eq:qflow}.

This set of nonlinear equations form the shared mathematical foundation of all three problems addressed in this work.

\subsection{Power Flow}
Power flow (PF) analysis determines the steady-state distribution of voltages and power throughout the network for a given set of generation and load conditions. Specifically, it computes the voltage magnitude and phase angle at every bus and the active and reactive power flows on all transmission branches.
 
PF analysis spans across a wide range of use cases, involving different time scales, like network planning and real-time operation, where it verifies that line loadings and bus voltages remain within security limits. \gls{pf} is governed by a nonlinear system of equations,~\eqref{eq:pflow}-\eqref{eq:qflow}, which must be solved simultaneously for all buses.

The industry-standard algorithm is the NR method, which linearizes the system around an initial operating point, conventionally a \emph{flat start} ($\abs{V} = 1$\,p.u., $\delta = 0$), and iteratively refines the state vector by solving a linear correction system governed by the Jacobian matrix $\mathbf{J}$. Convergence is typically achieved in five iterations under normal operating conditions \cite{tinney_power_1967}. The principal computational bottleneck is the formation and sparse factorization of $\mathbf{J}$ at each iteration, an operation whose cost scales super-linearly with system size.
 
This computational burden provides the primary motivation for the GNN-based approach proposed in this paper: once trained, the network approximates the mapping defined by \eqref{eq:pflow}-\eqref{eq:qflow} in a single forward pass, retaining topological awareness without repeated matrix factorizations or the sensitivity to initial conditions that can cause NR to diverge under stressed network conditions \cite{tinney_power_1967}.  
\subsection{Optimal Power Flow}
Optimal Power Flow is a mathematical optimization problem that determines the optimal operating state of an electric power system. While conventional \gls{pf} analysis simply calculates the steady-state conditions for a given set of generation and load values, OPF finds the best generation dispatch, voltage settings, transformer tap positions, and other control variables to optimize a specified objective while enforcing the full set of physical and operational constraints.

First formulated by Carpentier in \cite{carpentier_1962}, \gls{opf} has become a fundamental tool in power system operation. The problem has evolved from simple economic dispatch considerations to complex multi-objective optimization problems that consider environmental factors, market operations, and the integration of renewable energy sources.

The general OPF problem is formulated as: 
\begin{align}
  \min_{\mathbf{x},\mathbf{u}} \quad & f(\mathbf{x},\mathbf{u}) \label{eq:opf_obj}\\
  \text{s.t.} \quad & \mathbf{g}(\mathbf{x},\mathbf{u}) = \mathbf{0} \label{eq:opf_eq}\\
                    & \mathbf{h}(\mathbf{x},\mathbf{u}) \leq \mathbf{0}, \label{eq:opf_ineq}
\end{align}
   where $\mathbf{x} \in \mathbb{R}^{2n}$ is the state vector (voltage
magnitudes and angles), $\mathbf{u}$ is the vector of control
variables, $f(\mathbf{x},\mathbf{u})$ is the scalar objective. The equality constraint $g(\mathbf{x}, \mathbf{u}) = 0$ enforces nodal power balance at each bus, as defined by the \gls{kcl}~(\ref{eq:pbalance}) and (\ref{eq:qbalance}), while the inequality constraints $h(\mathbf{x}, \mathbf{u}) \leq 0$ capture physical and operational bounds, such as generator active and reactive power limits, nodal voltage bounds, and branch thermal capacity constraints.

\subsection{State Estimation}

\gls{se} reconstructs the complete operating state of
the network, the full set of voltage magnitudes and phase
angles, from a limited and inherently noisy set of real-time
measurements. It addresses three practical challenges that arise in
operational grids:
\begin{itemize}[leftmargin=*]
  \item \textit{Incomplete observability:} Field instrumentation does
    not cover every bus and branch; the state at unmonitored locations
    must therefore be inferred from the available data.
  \item \textit{Measurement noise:} Readings from current and voltage
    transformers are corrupted by instrument error. SE filters these disturbances to provide the best
    statistical estimate of the true state.
  \item \textit{Bad data detection:} SE identifies and rejects gross
    measurement errors, caused by meter faults or data transmission
    failures, that would otherwise corrupt the state estimate.
\end{itemize}

The most widely used SE method is the \gls{wls} estimator, which minimizes the weighted sum of squared 
measurement residuals, alternative methods include the
Least Absolute Value (LAV) estimator, and Kalman filtering, applicable to dynamic state
tracking.

The operational SE workflow comprises four stages: (i)~real-time
measurement collection and validation; (ii)~observability analysis
to confirm that the available measurements uniquely determine the
state; (iii)~state computation via \gls{wls} or an equivalent method; and (iv)~bad-data detection and identification through normalized residual testing.
 
The output of SE, a clean, complete state vector, feeds all higher-level grid-management functions: SCADA,  Energy Management Systems (EMS), OPF solvers, security assessment, automatic generation
control (AGC), economic dispatch, and contingency analysis. Any degradation in SE accuracy therefore propagates directly into all downstream applications \cite{monticelli_state_1999}.

\section{Graph Neural Network Background}
\label{sec:gnn background}

Power systems are inherently non-Euclidean: their behavior is governed by the admittance matrix and the adjacency structure of buses and lines/transformers, not by the Euclidean distance between nodes. \glspl{gnn} overcome these limitations by operating directly on the graph structure through an iterative \emph{message-passing} mechanism \cite{gilmer2017neuralmessagepassingquantum}, which updates node features by aggregating information from their immediate neighborhood. Each bus (node) updates its latent representation by aggregating information from its immediate electrical neighbors, closely mirroring the way
\gls{kcl} propagates information across the physical network. Because the same learned parameters are applied at every node and edge regardless of graph size or configuration, GNNs are permutation-invariant and can generalize across different topologies without retraining, properties that make them uniquely appropriate for the unified multi-task solver proposed in this work.

A standard GNN layer is shown at Fig.~\ref{fig:gnn_resgated}, where ${h_v}$ is the latent embedding of node $v$ and $\mathcal{N}(i)$ denotes the set of adjacent nodes. 

Early variants such as GCN and GAT demonstrated the viability of this paradigm for power system problems, but they suffer from limited discriminative power: it has been shown that these architectures cannot distinguish between certain non-isomorphic graph structures~\cite{xu_how_2019}, meaning different electrical networks. In power grids, where even a single line trip can substantially alter the operating state, this limitation is particularly consequential.

\subsection{ResGated GCN}

Introduced by Bresson and Laurent \cite{bresson_residual_2018} (2018), the \gls{resgated gcn} addresses the limitations of early graph neural network paradigms. While traditional Graph Recurrent Networks (GRNNs) offered complex memory but suffered from computational overhead, and early GCNs \cite{bruna_spectral_2014} provided speed but lacked depth scalability, \gls{resgated gcn} combines dynamic gating and residual connections to enable deep, efficient graph learning.

\begin{figure}[hbt]
    \centering    
    \begin{adjustbox}{width=1\width}
    \includegraphics[width=0.8\linewidth]{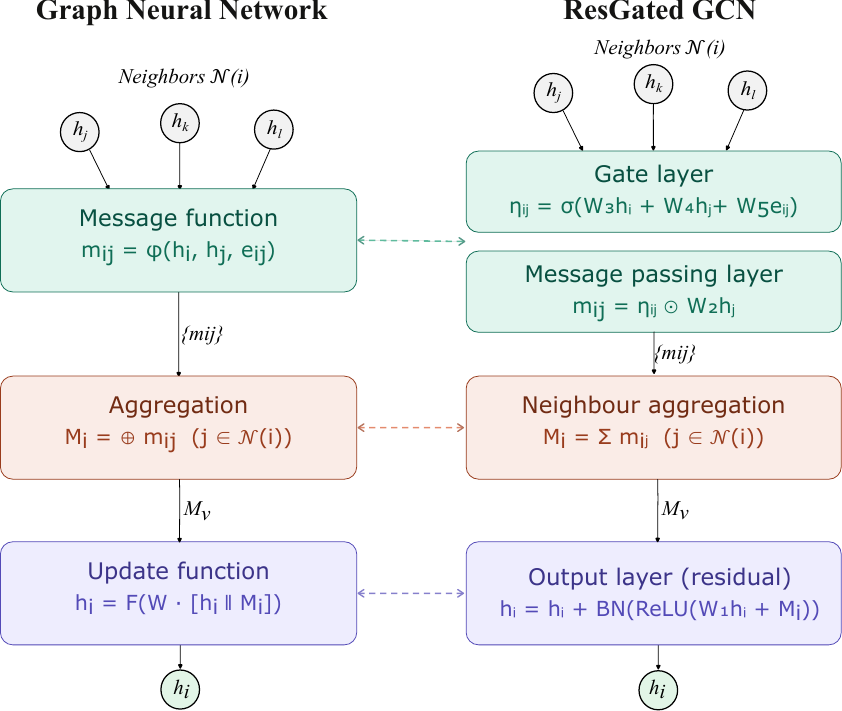}
    \end{adjustbox}
    \caption{GNN and ResGated GCN structure comparison.}
    \label{fig:gnn_resgated}
\end{figure}

The ResGated GCN layer updates a node's features by filtering neighbor messages through a dynamic, learned gate. The layer transformation is defined as (Fig.\ref{fig:gnn_resgated}), where the equation is extended to incorporate edge attributes, which is expressed as:\begin{center}
\footnotesize 
\begin{multline}
    h_{i}^{(l+1)} = h_{i}^{(l)} + \text{ReLU}\Bigg(\text{BN}\Bigg(W_{1}^{l}h_{i}^{(l)} 
    + \sum_{j \in \mathcal{N}(i)} \eta_{ij}^{l} \odot \left(W_{2}^{l}h_{j}^{(l)}\right)\Bigg)\Bigg)
    \label{eq:edge_mp}
 \end{multline}
\end{center}
\begin{equation}
    \eta_{ij}^{l} = \sigma\left(W_{3}^{l}h_{i}^{(l)} + W_{4}^{l}h_{j}^{(l)} + W_{e}^{l}e_{ij}\right) \label{eq:edge_gate}
\end{equation}
Let $h_i \in \mathbb{R}^d$ and $h_j \in \mathbb{R}^d$ denote the input features of the target node $i$ and a neighbouring node $j$, respectively, where $h_i$ also serves as a linear residual connection. The edge between them is characterised by $e_{ij} = [R_{ij}, X_{ij}, B_{ij}]^T$, which encodes the physical transmission line parameters, projected by the learnable matrix $W_e^\ell \in \mathbb{R}^{d' \times d_e}$. Four learnable weight matrices govern the layer transformations: $W_1^\ell \in \mathbb{R}^{d' \times d}$ applies a self-loop transform to the centre node $h_i$, while $W_2^\ell \in \mathbb{R}^{d' \times d}$ projects the neighbouring features $h_j$. The gating mechanism is controlled by $W_3^\ell, W_4^\ell \in \mathbb{R}^{d' \times d}$, which project the local topological states of $h_i$ and $h_j$, respectively, to compute the edge gate importance. The resulting gating coefficient $\eta_{ij}$, where $\sigma$ is the sigmoid activation, where $\sigma(x) = 1/(1+e^{-x})$, constraining values smoothly to $[0, 1]$, acts as an element-wise filter, applied via the Hadamard product $\odot$, that tracks the semantic relationship between connected nodes and controls information flow from each neighbour $j \in \mathcal{N}(i)$. Finally, batch normalisation $\mathrm{BN}(\cdot)$ and ReLU function, where $ReLU(x) = max(0, x)$ are applied to the aggregated output before the residual addition.

The main innovations and advantages of ResGated GCN are:

\begin{itemize}[leftmargin=*]
    \item \textit{Vector Edge Gating:} Unlike scalar attention (e.g., GAT), the gate $\boldsymbol{\eta}_{ij}$ is a vector of the same dimension as the hidden features, allowing independent modulation per dimension.
    \item \textit{Residual Connections:} The inclusion of a hard identity shortcut ($+ \mathbf{h}_{i}^{\ell}$) prevents vanishing gradients, enabling stable training of deeper architectures.
    \item \textit{Dual Self-Connection:} The model applies both an explicit linear projection ($\mathbf{W}_{1}\mathbf{h}_{i}$) within the aggregation and a hard residual addition after the non-linearity.
    \item \textit{Direction-Dependent Message Weighting:} By incorporating both endpoint features and explicit edge attributes $e_{ij}$, the network captures the full heterogeneity of transmission line parameters within the learned gate.
\end{itemize}

\subsection{Heterogeneous Graph Neural Networks for Power Systems}

Standard GNNs are \emph{homogeneous}: all nodes and edges share the same feature space and are updated by identical transformation functions. This assumption is incompatible with the structure of power systems, where buses represent physically distinct entities with different operational roles, boundary conditions, and sets of known and unknown variables.

\begin{figure}[htbp]
    \centering
    \subfloat[IEEE 9-bus system]{%
        \includegraphics[width=0.85\columnwidth]{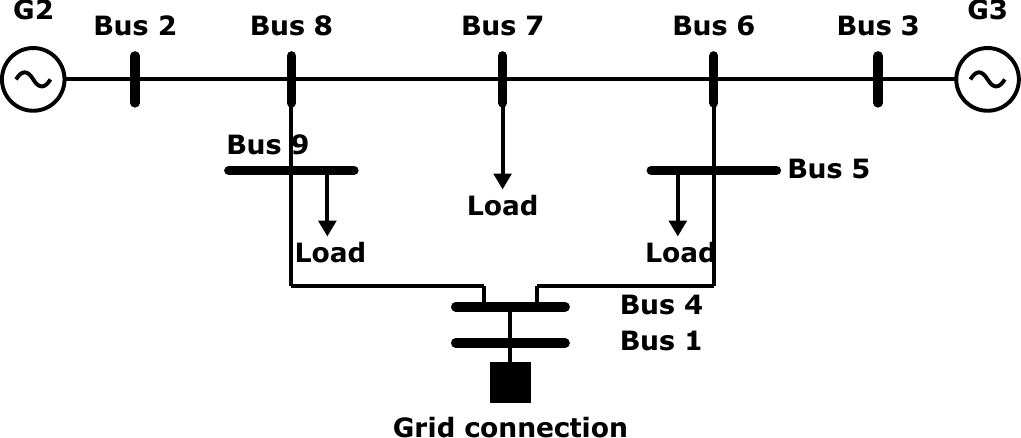}%
        \label{fig:sub_top}%
    }
    
    \subfloat[IEEE 9-bus graph]{%
        \includegraphics[width=0.85\columnwidth]{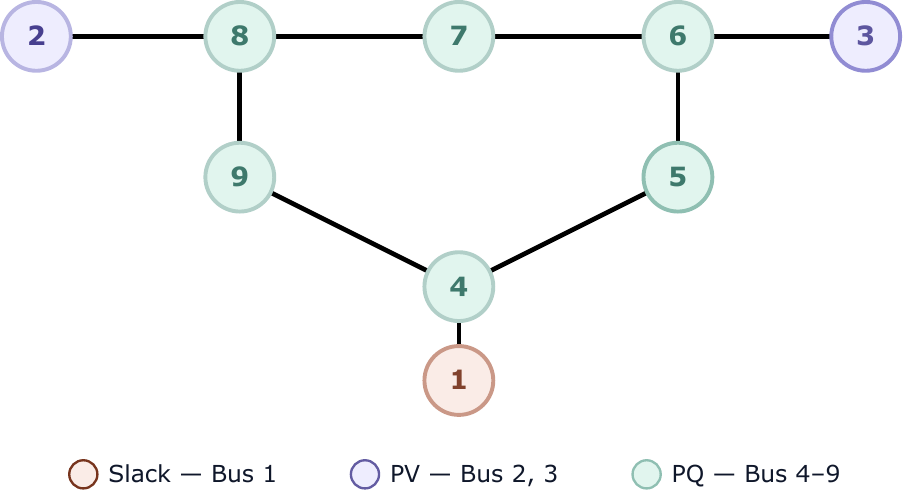}%
        \label{fig:sub_bottom}%
    }
    \caption{IEEE 9-bus system \cite{anderson1977power} represented as an electrical network and as a heterogeneous graph.}
    \label{fig:9-bus}
\end{figure}

In a homogeneous GNN, the distinct semantics of slack, PV, and PQ buses are flattened into a single feature representation, forcing the model to learn type-specific behavior purely from data rather than encoding it architecturally. This increases the effective learning difficulty and risks conflating physically distinct
operating modes. \glspl{hgnn} resolve this limitation by associating each node and edge type with its own learnable parameters and message-passing logic, motivated by two principal considerations:

\begin{itemize}[leftmargin=*]
    \item \textit{Bus-Type Specificity}: Electrical buses are defined by which variables are known and unknown. A \textit{slack} bus has a fixed voltage magnitude and angle ($V, \theta$); a \textit{PV} bus has fixed active power and voltage magnitude ($P, V$); and a \textit{PQ} bus has fixed active and reactive power ($P, Q$). \glspl{hgnn} allow the architecture to apply specific transformation matrices to each bus type, ensuring that the GNN respects these distinct boundary conditions during the feature aggregation process. An example based on the IEEE 9-bus is shown at Fig. \ref{fig:9-bus}
    \item \textit{Edge Physics}: Unlike social or citation networks, the edges in a power grid (transmission lines and transformers) carry critical physical parameters such as resistance ($R$), reactance ($X$), and shunt admittance ($B$), for transmission lines, or transformation relation and phase shift for transformers.
\end{itemize}

\section{Proposed Methodology}
\label{sec:proposed methodology}
\begin{figure}[htbp]
    \centering
    \begin{adjustbox}{width=0.70\columnwidth}
    \includegraphics[width=0.2\linewidth]{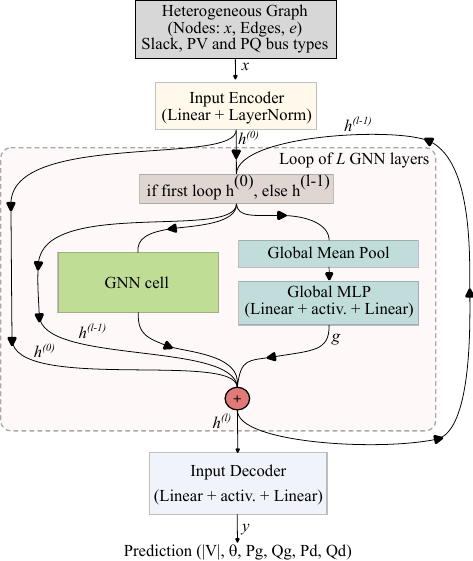}
    \end{adjustbox} 
    \caption{Architecture of the proposed unified Heterogeneous ResGated GCN.}
    \label{fig:gnn_architecture}
\end{figure}

The proposed architecture (Fig. \ref{fig:gnn_architecture}) is a deep \glspl{hgnn} designed to solve PF, OPF, and SE within a single unified architecture, based on the Encoder-Processor-Decoder architecture proposed by \cite{sanchez-gonzalez_learning_2020}, extended to heterogeneous graphs. By treating buses of different types (\textit{slack, PV, PQ}) as distinct node sets $\mathcal{V}_k$ with independent encoder and decoder parameters, the model enforces type-specific boundary conditions for each task while sharing a common physics-informed message-passing backbone across all three problems.

\subsection{Architectural Components}

\subsubsection{Node Encoding and Feature Transformation}
To handle the heterogeneous and task-dependent input features across bus types, we employ a bank of type-specific linear encoders. For each node type
$k \in \{\mathrm{slack}, \mathrm{pv}, \mathrm{pq}\}$, the raw feature vector $\mathbf{x}_i \in \mathbb{R}^{d_{\mathrm{in}}}$ is projected into a shared latent space $\mathbb{R}^{d_h}$, where $d_h$ is the hidden size, via a
type-specific affine transformation followed by Layer
Normalization:
\begin{equation}
  h_i^{(0)} = \mathrm{LayerNorm}\!\left(\mathbf{W}_k\,\mathbf{x}_i + \mathbf{b}_k\right).
  \label{eq:encoding}
\end{equation}
where $\mathbf{W}_k$ are separate weight matrices for each bus type ensures that the model distinguishes the fixed-voltage boundary conditions of a slack bus from the power-injection inputs of a PQ bus from the first layer onward, and $b_k$ is the bias term. Layer Normalization stabilizes training by preventing feature-scale disparities from propagating through the deep stack.

\subsubsection{Heterogeneous Message Passing}

The core of the architecture consists of $L$ stacked \gls{hgnn} layers, each applying separate transformation matrices to each directed relation type (e.g.,\textit{ slack-to-PQ, PV-to-PQ}). At each layer, the edge attribute vector $\mathbf{e}_{ij}$, encoding the branch impedance parameters $(R, X)$ and shunt admittance $(B)$, is incorporated into message passing according to~\eqref{eq:edge_mp}-\eqref{eq:edge_gate}. 
By applying the non-linear activation to the sum of the neighbor embedding and the projected edge attributes before aggregation, the model learns the nonlinear relationships that Ohm's and \gls{kcl} impose on nodal voltages and power injections.
 
\subsubsection{Global Context and Residual Integration}
OPF requires awareness of global system constraints, such as total generation cost and network-wide reactive power balance, that local message passing alone cannot efficiently propagate in deep architectures. A global context vector, $\mathbf{g}$, is therefore computed at each layer by mean-pooling
the current node embeddings across all bus types, $k$, and passing them through a shared MLP:
\begin{equation}
\mathbf{g} = \text{MLP}_{global}\left(\frac{1}{\sum_k|\mathcal{V}_k|}
\sum_{k \in \text{Types}} \sum_{i \in \mathcal{V}_k} 
\mathbf{h}_i^{(l)}\right)
\end{equation}
where $|\mathcal{V}| = \sum_{k}|\mathcal{V}_k|$ denotes the total number of nodes across all bus types, so that the mean-pooling normalizes over the full graph regardless of the type partition.

The global vector is broadcast back to all nodes and combined with the per-node update via a multi-path residual connection:
\begin{equation}
    \mathbf{h}_i^{(l)} = \mathbf{h}_i^{(l-1)} + 
    \text{SiLU}\!\left(\mathbf{h}_{i,\text{new}}^{(l)}\right) + 
    \mathbf{g} + \mathbf{h}_i^{(0)}
    \label{eq:residual}
\end{equation}
where $\text{SiLU}(x) = x/(1+e^{-x})$ is applied element-wise. Note that $\text{SiLU}(\cdot)$ is distinct from the sigmoid gate $\sigma(\cdot)$ in \eqref{eq:edge_gate}, which is reserved exclusively for the ResGated GCN gating mechanism throughout this paper. The skip connection to the initial encoding $h_i^{(0)}$ is particularly important: it ensures that the task-specific and bus-type boundary conditions encoded in the input projection remain influential throughout the full network depth, preventing the model from losing track of the physical constraints of the problem as depth increases, due to oversmoothing, a well-known problem in \gls{gnn}s.

\subsubsection{Multi-Head Decoding}
The final node embeddings $\mathbf{h}_i^{(L)}$ are decoded by task using variable-specific output heads, providing greater expressivity for quantities of different physical nature — notably voltage magnitude and voltage angle, which span different numerical ranges. The output biases of voltage magnitude and angle are initialized to physically meaningful values: $1.0$\,p.u. for voltage magnitude and $0.0$\,rad for voltage angle, corresponding to the flat-start initialization used in classical iterative solvers. This physics-informed initialization places the initial output distribution in a physically plausible region, reducing the number of training epochs required to reach the target accuracy regime.

\subsection{Unified Multi-Task Training Objective}
\label{subsec:loss}
A central challenge in training a unified solver is the
heterogeneous character of the three tasks: PF and SE are regression problems with smooth loss profile, while OPF is a constrained optimization problem in which constraint satisfaction is as operationally critical as prediction accuracy. To address this, we employ  the mean squared error between predicted and ground-truth state variables, providing the primary gradient signal for all three tasks.    
    \begin{equation}
        MSE = \frac{1}{n} \sum_{i=1}^{n} (y_i - \hat{y}_i)^2
    \end{equation}
    where $n$ is the total number of observations, $y_i$ represents the actual (observed) values and $\hat{y}_i$ represents the predicted (simulated) values.

For each task, $\mathrm{MSE}$ is computed as a sum of per-variable terms, applied only to the state variables that are outputs for the corresponding bus type and task:
\begin{equation}
\mathcal{L}_{MSE} = \sum_{k \in \mathcal{V}} (\mathcal{L}_{Vm} + \mathcal{L}_{Va} + \mathcal{L}_{Pg} + \mathcal{L}_{Qg})
\end{equation}
For example, during a PF forward pass, $\mathcal{L}_{V_m}$ is applied only to PQ buses (for which $\abs{V}$ is unknown), while $\mathcal{L}_{V_a}$ is applied to both PQ and PV buses. This task-conditioned masking ensures that the model is penalized only for quantities it is responsible for predicting, preventing
spurious gradients from degrading the shared representation.

\subsection{Data Generation and Pre-processing}
The reliability of a GNN-based solver depends critically on the diversity and physical consistency of its training data. A model trained on a narrow distribution of operating conditions will
exhibit poor generalization to scenarios outside that distribution, a phenomenon known as \emph{out-of-distribution} (OOD) failure, which can manifest at the feature level (input values outside the training range) or at the topological level (changes in network connectivity, as arise in N-1 contingency
analysis). To mitigate this risk, a comprehensive data generation pipeline was developed using the \textit{PandaPower}\cite{noauthor_pandapower_nodate} and \textit{VeraGrid}\cite{vera_sanpenveragrid_2026} Python packages.
 
The following randomization strategy was applied to four different dataset scenarios, to maximize the diversity of operating conditions (Table \ref{tab:dataset}):

\begin{itemize}[leftmargin=*]
  \item \textit{Load variability:} Active and reactive power demands ($P_L$, $Q_L$) were sampled uniformly over a defined range around the nominal values. This wide range, spanning both light-load off-peak and heavily stressed peak conditions, ensures that the model encounters the full extent of the feasible operating region, including near-voltage-collapse scenarios.
  \item \textit{Generator and line parameter perturbation:} Generator voltage setpoints ($V_{\mathrm{gen}}$) and transmission line impedance ($Z_{ij}$) were varied within $\pm 5\%$ of their nominal values, simulating set-point uncertainty and parametric drift in aging infrastructure.
  \item \textit{Topological augmentation:} The network topology was modified by randomly removing existing branches or adding new ones. This structural diversity trains the \gls{gnn} backbone to recognize that the governing physics is encoded in the admittance matrix and line parameters, not in a fixed node ordering, improving robustness to grid reconfigurations and N-1 contingencies.
  \item \textit{Measurement noise and masking for SE:} State Estimation samples are further processed to replicate the conditions of a real-time operational environment. Each measurement is corrupted by additive zero-mean Gaussian noise, with a standard deviation drawn uniformly at random from the interval $[1\%, 2\%]$ of the corresponding measurement magnitude, simulating the accuracy class of typical current and voltage instrument transformers. In addition, 5\% of measurements are masked (set to missing) per sample, with the masked subset selected uniformly at random across all available measurements, simulating partial observability due to communication failures or instrumentation gaps. These perturbations are applied exclusively to SE samples, ensuring that the GNN learns to reconstruct the full system state from incomplete and noisy observations, consistent with the operational role of a state estimator.
\end{itemize}

To facilitate multi-task learning within a single shared model, we augmented the node feature matrix with a one-hot encoded task-indicator vector $\tau \in \{0, 1\}^3$:
\begin{equation}
    \tau_s = 
    \begin{cases} 
    [1, 0, 0]^T & \text{if } s \in \text{Power Flow} \\
    [0, 1, 0]^T & \text{if } s \in \text{OPF} \\
    [0, 0, 1]^T & \text{if } s \in \text{State Estimation}
    \end{cases}
    \label{eq:one-hot-task}
\end{equation}
This allows the heterogeneous encoders to condition the initial embeddings on the specific requirements of the objective task.

\begin{table}[htbp]
\centering
\caption{Scenario Definitions.}
\label{tab:dataset}
\setlength{\tabcolsep}{4pt} 
  \begin{tabular}{ll}
    \toprule
    \textbf{Dataset Name} & \textbf{Power Ranges}  \\
    \midrule
    \textit{narrow} & 50\% to 150\% nominal powers \\
    \textit{mid} & 20\% to 180\% nominal powers \\
    \textit{wide} & 0\% to 200\% nominal powers \\
    \textit{high-topo} & 180\% to 250\% nominal powers and \\ & 40\% samples with topological changes \\    
    \bottomrule
  \end{tabular}
\end{table}

\begin{table}[htbp]
\centering
\caption{Number of Rejected Samples Due to Infeasibility or Solver Non-Convergence During Data Generation for the IEEE 118-Bus Network.}
\label{tab:dataset-rejected}
\begin{tabular}{lcccccccc}
\toprule
\textbf{118-bus case}  & narrow & mid & wide & \makecell{high-topo} \\
\midrule
PF    & 0 & 0 & 5  & 542\\
OPF   & 0 & 0 & 1  & 546\\
SE    & 0 & 0 & 9  & 3045\\
\bottomrule
\end{tabular}

\end{table}

Table~\ref{tab:dataset} defines the four load scenarios, 
from the most conservative (\textit{narrow}) to the most 
heavily stressed (\textit{high-topo}), the latter including 
40\% of samples with random topological modifications. It is worth noting that the \textit{high-topo} scenario operates near the boundary of the feasibility region: in some cases, more than ten candidate samples had to be discarded due to solver non-convergence before a feasible operating point was found (Table \ref{tab:dataset-rejected}).

\subsection{Evaluation and Generalization Metrics}
\gls{mse} serves as the primary training objective but is an insufficient standalone evaluation metric for power system applications, where worst-case performance and constraint satisfaction are operationally critical. The model is therefore evaluated using the following complementary metrics, as proposed in \cite{jadhav_enhancing_2025}:

\begin{itemize}[leftmargin=*]
    \item \textit{Normalized Root  Mean Square Error (NRMSE):} Indicator based on \gls{mse}, which is normalized by the range of the ground truth output, allowing a correct comparison between different magnitudes.
 \begin{equation}
        \text{NRMSE} = \frac{\text{RMSE}}{y_{\max} - y_{\min}} = \frac{\sqrt{\frac{1}{n} \displaystyle\sum_{i=1}^{n} (y_i - \hat{y}_i)^2}}{y_{\max} - y_{\min}}
        \end{equation}
    where $y_{\max}$ and $y_{\min}$ are the maximum and minimum actual values in the dataset, respectively.
  \item \textit{Coefficient of determination ($R^2$):}
    \begin{equation}
      R^2 = 1 - \frac{\sum_i (y_i - \hat{y}_i)^2}
                     {\sum_i (y_i - \bar{y})^2},
      \label{eq:r2}
    \end{equation}
    measures the proportion of variance in the target variables explained by the model. Values close to unity confirm a high goodness-of-fit across the full operating range.
 
  \item \textit{Maximum Absolute Error (MaxAE):}
    \begin{equation}
      \mathrm{MaxAE} = \max_i \abs{y_i - \hat{y}_i},
      \label{eq:maxae}
    \end{equation}
    identifies the single largest prediction error across all buses and test samples. In operational contexts, worst-case accuracy is often more consequential than average accuracy,
    since a single large voltage or power error can violate security limits.
 
\end{itemize}

\section{Results and Discussion}
\label{sec:results}

\subsection{Simulation Setup}

A systematic study was conducted along two axes. First, a hyperparameter search varied the number of GNN layers over {2, 4, 6} and the hidden dimension over {32, 64, 128, 256}. Second, an ablation study individually removed the residual skip connection $h^{(0)}$, global mean pooling, and Layer Normalization  (\texttt{NormLayer}) to quantify the contribution of each architectural component.

The selected configuration was then evaluated on both the IEEE 14-bus and 118-bus test systems, using a combined dataset of 3000 samples, 1000 samples for each problem (PF/OPF/SE ), drawn from the \textit{mid} scenario and partitioned into 80\% training, 10\% validation, and 10\% test sets. To assess out-of-distribution generalization, the trained model on the \textit{mid} scenario was further evaluated on the \textit{narrow}, \textit{wide}, and \textit{high-topo} datasets, which include unseen loading conditions and network topologies, where existing lines can be removed or new lines added.

\subsection{Results}

Table~\ref{tab:comparison} and Figs.~\ref{fig:OPF-SE-results}-\ref{fig:PF-results} present the quantitative results of the hyperparameter search, ablation study, and generalization evaluation. 

Fig.~3 compares the average NRMSE, computed as the mean across all predicted quantities (\textit{$V_{mag}$}, \textit{$V_{ang}$}, \textit{$P_{gen}$}, \textit{$Q_{gen}$}, \textit{$P_{d}$} and \textit{$Q_{d}$}), for seven GNN cell variants: GAT, Transformer, ResGated GCN, GraphSAGE, GPS, GiNE, and MPNN. All variants are evaluated within the proposed heterogeneous architecture, using a 300 samples dataset, for the 118-bus network under the \textit{mid} scenario. ResGated GCN achieves the lowest average NRMSE and is therefore selected as the computational cell for all subsequent experiments.

\begin{figure}[ht]
    \centering
    \includegraphics[width=1\linewidth]{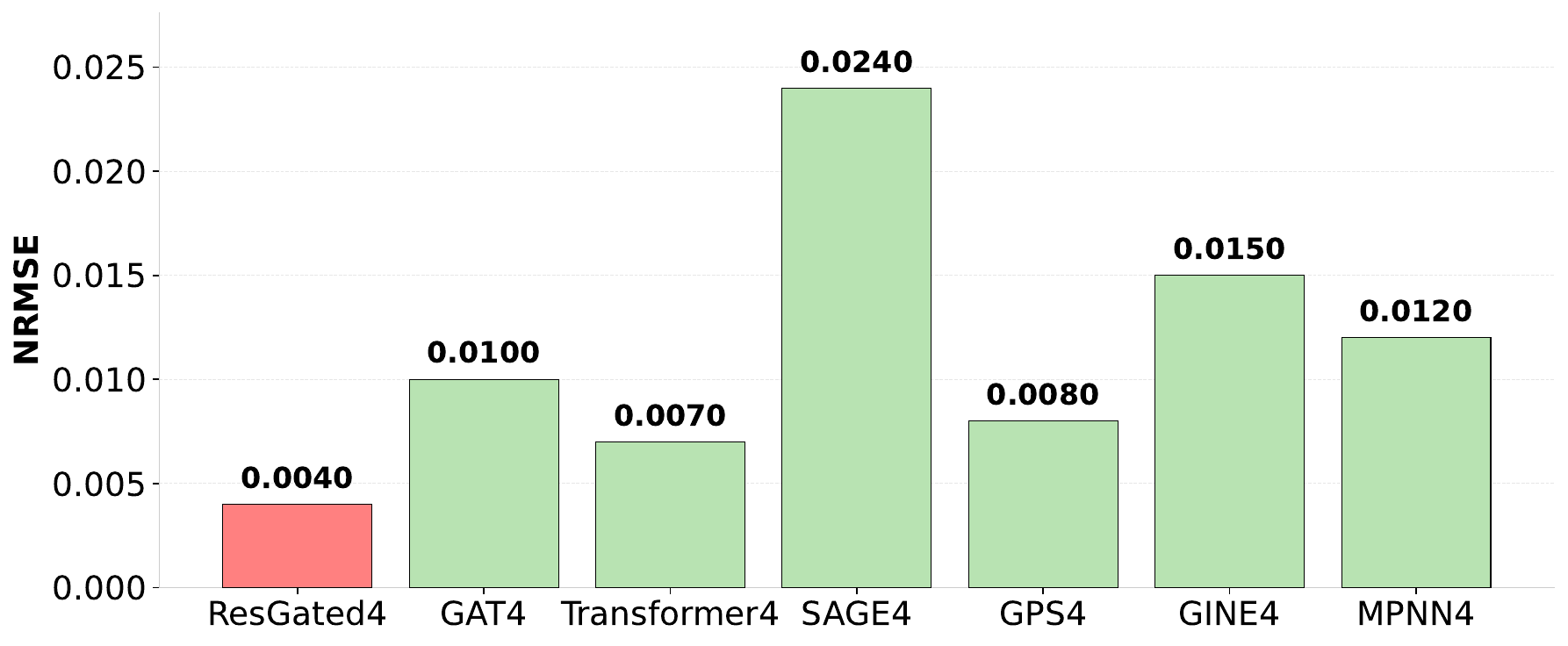}
    \caption{GNN cell comparison showing average NRMSE.}
    \label{fig:GNNcomparison}
\end{figure}

\begin{figure*}[t]
    \centering
    \includegraphics[width=0.99\linewidth]{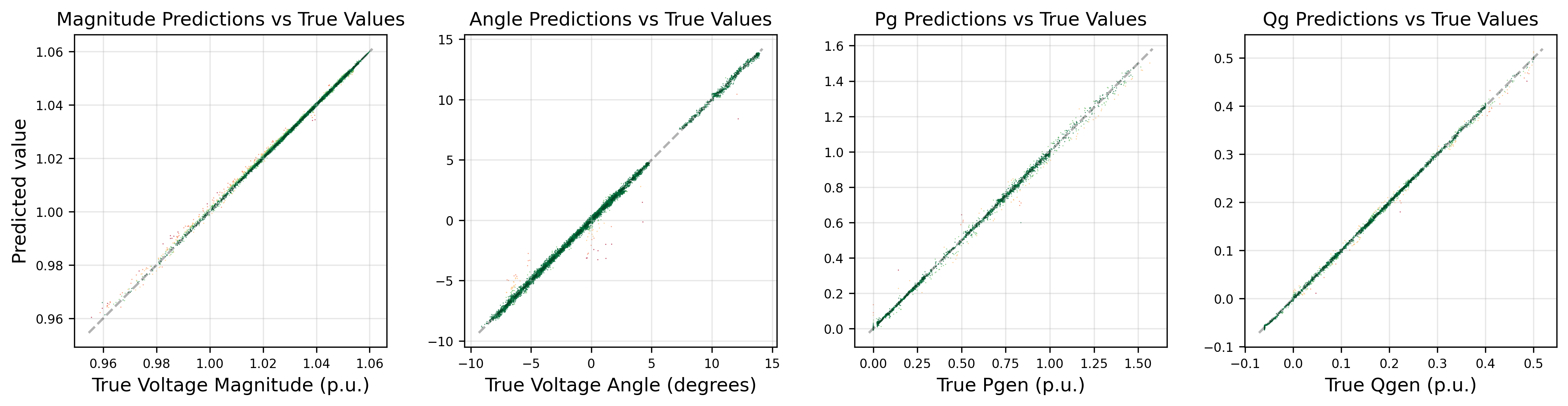}
    \includegraphics[width=1\linewidth]{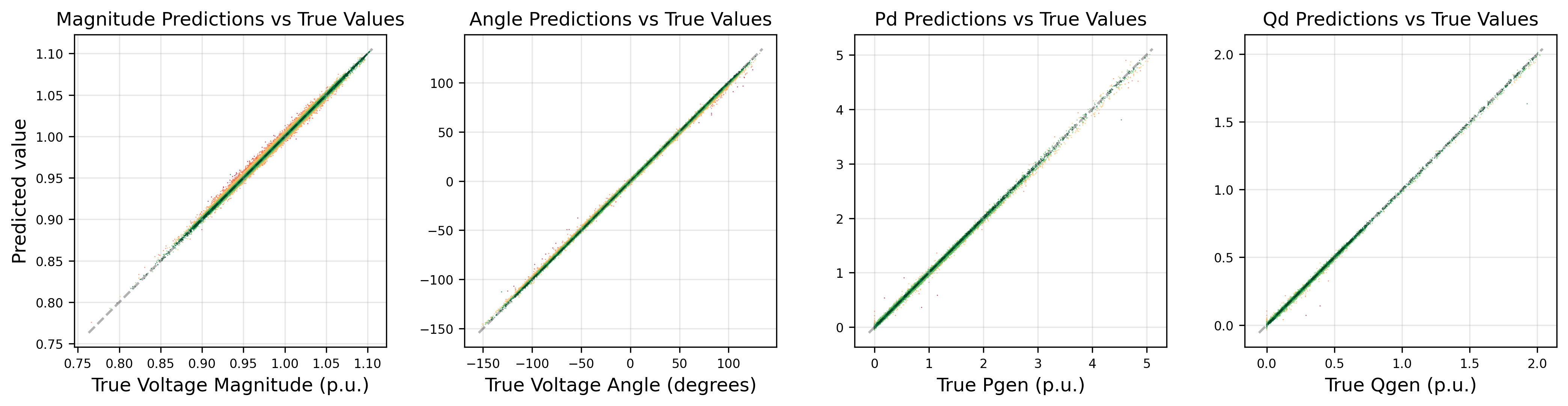}
    \caption{Prediction vs. ground truth scatter plots for the OPF task (IEEE 14-bus network, top row) and the SE task (IEEE 118-bus network, bottom row), both trained and evaluated on their respective \textit{mid} scenario datasets.}
    \label{fig:OPF-SE-results}
\end{figure*}

Table~\ref{tab:comparison} summarizes the hyperparameter search and ablation results for the 118-bus network, trained on the \textit{mid} scenario and tested on both the \textit{mid} and \textit{high-topo} scenarios. Inference times are reported relative to the fastest configuration to remove hardware-dependent bias. Fig.~\ref{fig:AverageNRSMSE-Ablation} complements the table by showing the average NRMSE graphically for each ablated configuration. Fig~\ref{fig:OPF-SE-results} shows two prediction cases, which exemplify two problems, OPF and SE, using two different networks, IEEE 14-bus and IEEE 118-bus network, both trained and tested on the \textit{mid} scenario. Fig.~\ref{fig:training_loss} shows the training and validation loss evolution over 250 epochs for the IEEE 118-bus network, trained on the \textit{mid} scenario, the loss is plotted on a logarithmic scale, encompassing PF, OPF, and SE simultaneously. Both curves converge smoothly without signs of overfitting. Fig.~\ref{fig:PF-results} shows PF results for the 118-bus network evaluated across all four scenarios (\textit{narrow}, \textit{mid}, \textit{wide}, \textit{high-topo}), confirming that prediction errors remain small and tightly distributed around zero even under conditions not seen during training.

\begin{figure}
    \centering
    \includegraphics[width=0.95\linewidth]{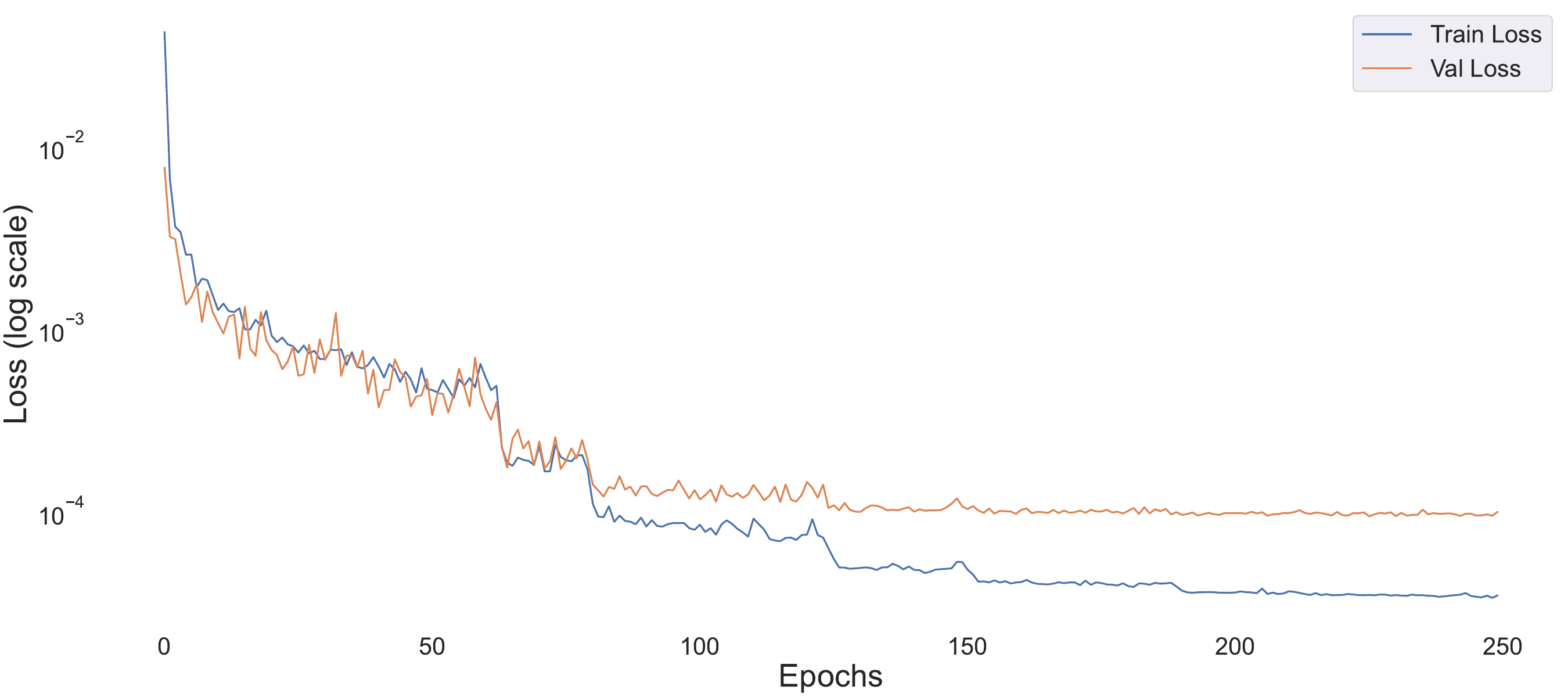}
    \caption{Training and validation loss evolution over 250 epochs for the IEEE 118-bus network, trained on the \textit{mid} scenario.}
    \label{fig:training_loss}
\end{figure}

\subsection{Discussion}

\begin{figure}
    \centering
    \includegraphics[width=1\linewidth]{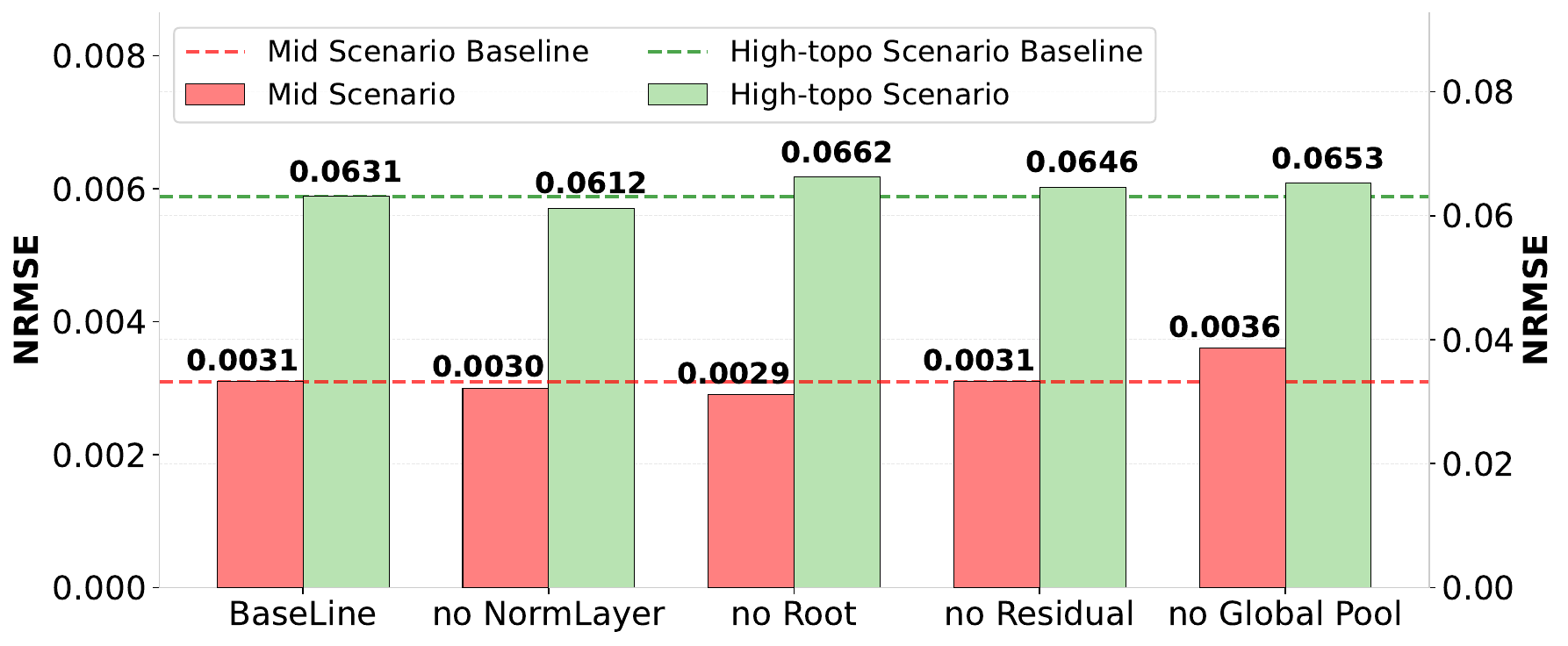}
    \caption{Ablation study results showing average NRMSE.}
    \label{fig:AverageNRSMSE-Ablation}
\end{figure}

\begin{table*}[ht]
\centering
\caption{Comparison of Prediction Metrics Across Configurations. Base Line: 4 GNN Layers and Hidden Dimension of 64.}
\label{tab:comparison}
\tiny
\setlength{\tabcolsep}{4pt} 
\begin{tabular}{lccccccccccc}
\toprule
 &inference & avg & \multicolumn{4}{c}{Vmag} & Vang & Pgen & Qgen & Pd & Qd \\
\cmidrule(lr){3-3} \cmidrule(lr){4-7} \cmidrule(lr){8-8} \cmidrule(lr){9-9} \cmidrule(lr){10-10} \cmidrule(lr){11-11} \cmidrule(lr){12-12}
Case & time & NRMSE & NRMSE & MAE & R2 & MSE & NRMSE & NRMSE & NRMSE & NRMSE & NRMSE \\
\midrule
\cellcolor{lightgray}{\textbf{Mid Scenario}}\\
Base line & $1$ & $\bm{0.0031}$ & $0.0056 \pm 0.000$ & $0.001$ & $0.998$ & $2.0587e-06$ & $0.0033 \pm 0.000$ & $0.0047 \pm 0.000$ & $0.0024 \pm 0.000$ & $0.0015 \pm 0.000$ & $0.0011 \pm 0.000$ \\
2 GNN layers & $0.8$ & $0.0034$ & $0.0058 \pm 0.000$ & $0.001$ & $0.998$ & $2.1972e-06$ & $0.0038 \pm 0.000$ & $0.0052 \pm 0.000$ & $0.0028 \pm 0.000$ & $0.0016 \pm 0.000$ & $0.0013 \pm 0.000$ \\
6 GNN layers & $1.73$ & \textcolor{ForestGreen}{$0.0030$} & $0.0052 \pm 0.001$ & $0.001$ & $0.999$ & $1.7727e-06$ & $0.0034 \pm 0.000$ & $0.0046 \pm 0.000$ & $0.0023 \pm 0.000$ & $0.0017 \pm 0.000$ & $0.0011 \pm 0.000$ \\
32  hidden dim.& $0.86$ & $0.0037$ & $0.0065 \pm 0.000$ & $0.001$ & $0.998$ & $2.7967e-06$ & $0.0041 \pm 0.000$ & $0.0053 \pm 0.000$ & $0.0028 \pm 0.000$ & $0.0020 \pm 0.000$ & $0.0015 \pm 0.000$ \\
128 hidden dim. & $1.2$ & \textcolor{ForestGreen}{$0.0029$} & $0.0050 \pm 0.000$ & $0.001$ & $0.999$ & $1.6324e-06$ & $0.0029 \pm 0.000$ & $0.0048 \pm 0.000$ & $0.0023 \pm 0.000$ & $0.0015 \pm 0.000$ & $0.0010 \pm 0.000$ \\
256 hidden dim. & $1.66$ & \textcolor{ForestGreen}{$0.0029$} & $0.0051 \pm 0.000$ & $0.001$ & $0.999$ & $1.6871e-06$ & $0.0029 \pm 0.000$ & $0.0048 \pm 0.000$ & $0.0023 \pm 0.000$ & $0.0015 \pm 0.000$ & $0.0010 \pm 0.000$ \\
no Global Pool & $0.93$ & $0.0036$ & $0.0076 \pm 0.000$ & $0.001$ & $0.997$ & $3.8088e-06$ & $0.0033 \pm 0.000$ & $0.0053 \pm 0.000$ & $0.0025 \pm 0.000$ & $0.0017 \pm 0.000$ & $0.0012 \pm 0.000$ \\
no Residual & $0.97$ & $0.0031$ & $0.0051 \pm 0.000$ & $0.001$ & $0.999$ & $1.7124e-06$ & $0.0031 \pm 0.000$ & $0.0050 \pm 0.000$ & $0.0026 \pm 0.000$ & $0.0016 \pm 0.000$ & $0.0009 \pm 0.000$ \\
no Root & $0.97$ & \textcolor{ForestGreen}{$0.0029$} & $0.0050 \pm 0.000$ & $0.001$ & $0.999$ & $1.6401e-06$ & $0.0030 \pm 0.000$ & $0.0047 \pm 0.000$ & $0.0024 \pm 0.000$ & $0.0015 \pm 0.000$ & $0.0010 \pm 0.000$ \\
no NormLayer & $0.98$ & \textcolor{ForestGreen}{$0.0030$} & $0.0053 \pm 0.000$ & $0.001$ & $0.999$ & $1.8427e-06$ & $0.0029 \pm 0.000$ & $0.0050 \pm 0.000$ & $0.0024 \pm 0.000$ & $0.0015 \pm 0.000$ & $0.0010 \pm 0.000$ \\
\cellcolor{lightgray}{\textbf{High-topo Scenario}}\\
Base line && $\bm{0.0631}$ & $0.0523 \pm 0.003$ & $0.007$ & $0.840$ & $1.6655e-04$ & $0.1088 \pm 0.003$ & $0.1026 \pm 0.004$ & $0.0719 \pm 0.002$ & $0.0218 \pm 0.002$ & $0.0209 \pm 0.002$ \\
2 GNN layers && \textcolor{ForestGreen}{$0.0613$} & $0.0519 \pm 0.004$ & $0.007$ & $0.842$ & $1.6435e-04$ & $0.1061 \pm 0.002$ & $0.1090 \pm 0.002$ & $0.0729 \pm 0.004$ & $0.0149 \pm 0.002$ & $0.0129 \pm 0.001$ \\
6 GNN layers && $0.0640$ & $0.0572 \pm 0.002$ & $0.008$ & $0.809$ & $1.9879e-04$ & $0.1036 \pm 0.008$ & $0.1075 \pm 0.003$ & $0.0696 \pm 0.003$ & $0.0233 \pm 0.000$ & $0.0228 \pm 0.002$ \\
32  hidden dim.&& $0.0635$ & $0.0548 \pm 0.002$ & $0.007$ & $0.824$ & $1.8273e-04$ & $0.1039 \pm 0.006$ & $0.1065 \pm 0.010$ & $0.0742 \pm 0.008$ & $0.0224 \pm 0.001$ & $0.0191 \pm 0.000$ \\
128 hidden dim.&& $0.0639$ & $0.0539 \pm 0.002$ & $0.007$ & $0.830$ & $1.7683e-04$ & $0.1153 \pm 0.004$ & $0.0987 \pm 0.002$ & $0.0709 \pm 0.002$ & $0.0214 \pm 0.002$ & $0.0229 \pm 0.001$ \\
256 hidden dim.&& $0.0640$ & $0.0525 \pm 0.001$ & $0.007$ & $0.839$ & $1.6731e-04$ & $0.1210 \pm 0.002$ & $0.0966 \pm 0.002$ & $0.0696 \pm 0.002$ & $0.0204 \pm 0.001$ & $0.0239 \pm 0.001$ \\
no Global Pool && $0.0653$ & $0.0504 \pm 0.004$ & $0.007$ & $0.851$ & $1.5503e-04$ & $0.1217 \pm 0.005$ & $0.1091 \pm 0.018$ & $0.0712 \pm 0.006$ & $0.0201 \pm 0.000$ & $0.0195 \pm 0.001$ \\
no Residual && $0.0646$ & $0.0557 \pm 0.004$ & $0.007$ & $0.818$ & $1.8895e-04$ & $0.1036 \pm 0.003$ & $0.1075 \pm 0.006$ & $0.0732 \pm 0.005$ & $0.0240 \pm 0.002$ & $0.0238 \pm 0.001$ \\
no Root && $0.0662$ & $0.0542 \pm 0.002$ & $0.007$ & $0.828$ & $1.7866e-04$ & $0.1130 \pm 0.002$ & $0.1090 \pm 0.004$ & $0.0741 \pm 0.003$ & $0.0229 \pm 0.002$ & $0.0239 \pm 0.001$ \\
no NormLayer && \textcolor{ForestGreen}{$0.0612$} & $0.0510 \pm 0.003$ & $0.007$ & $0.848$ & $1.5807e-04$ & $0.1117 \pm 0.004$ & $0.0997 \pm 0.016$ & $0.0715 \pm 0.005$ & $0.0154 \pm 0.001$ & $0.0177 \pm 0.002$ \\
\bottomrule
\end{tabular}

\end{table*}

\begin{figure*}[htbp]
    \centering
    \subfloat[\textit{narrow}]{%
        \includegraphics[width=0.24\textwidth]{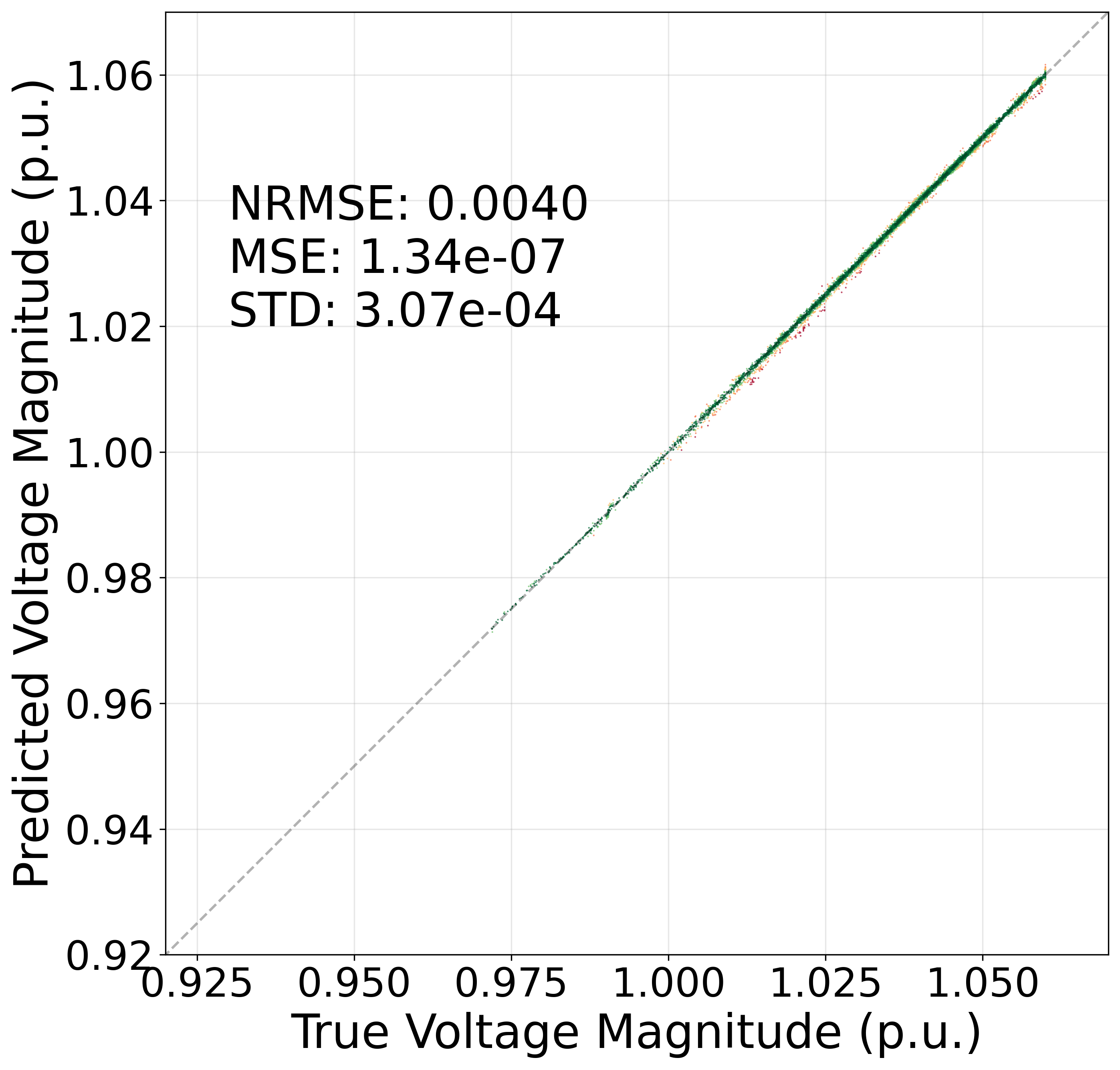}%
        \label{fig:sub1}%
    }%
    \hfill
    \subfloat[\textit{mid}]{%
        \includegraphics[width=0.24\textwidth]{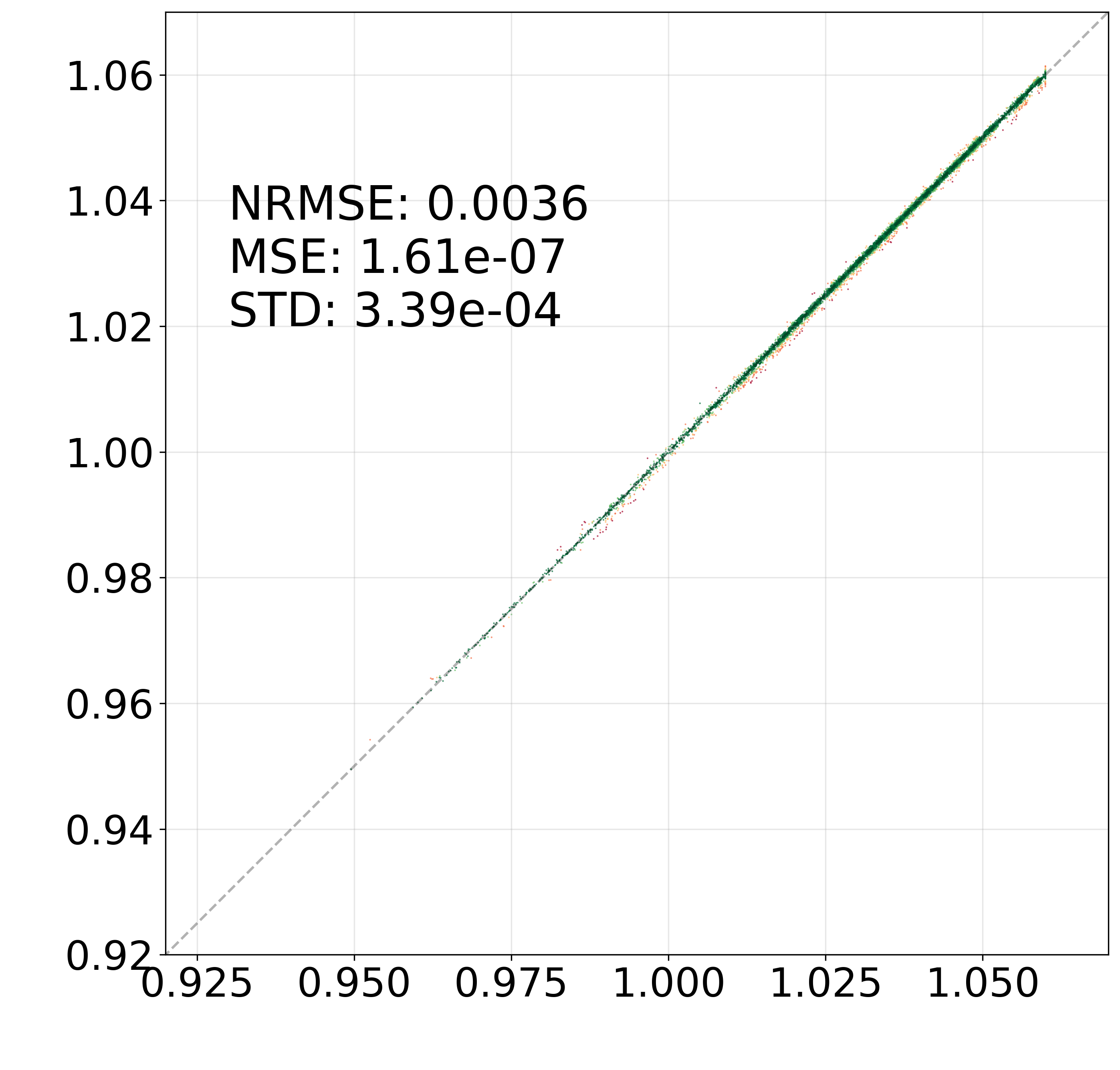}%
        \label{fig:sub2}%
    }%
    \hfill
    \subfloat[\textit{wide}]{%
        \includegraphics[width=0.24\textwidth]{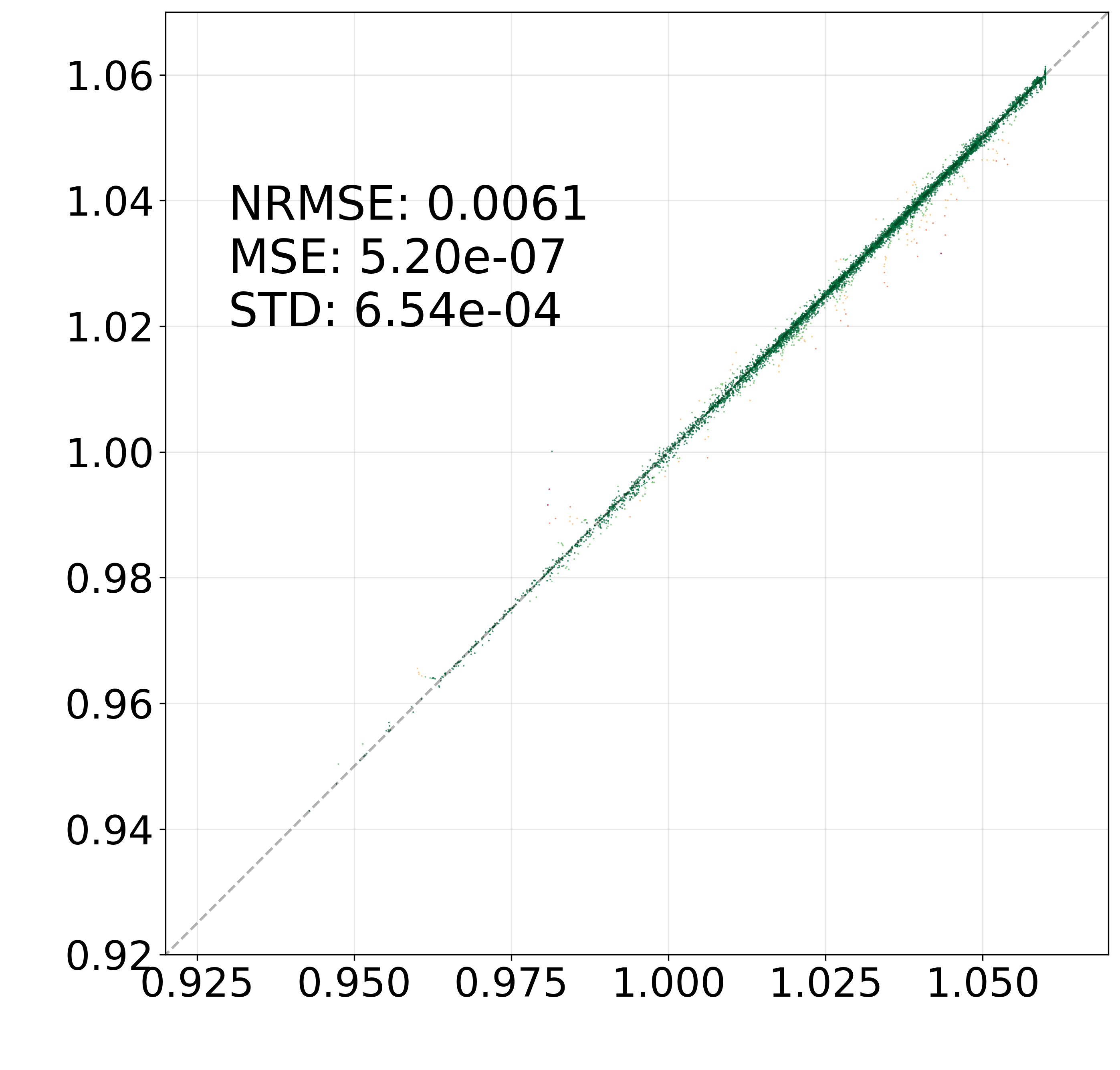}%
        \label{fig:sub3}%
    }%
    \hfill
    \subfloat[\textit{high-topo}]{%
        \includegraphics[width=0.24\textwidth]{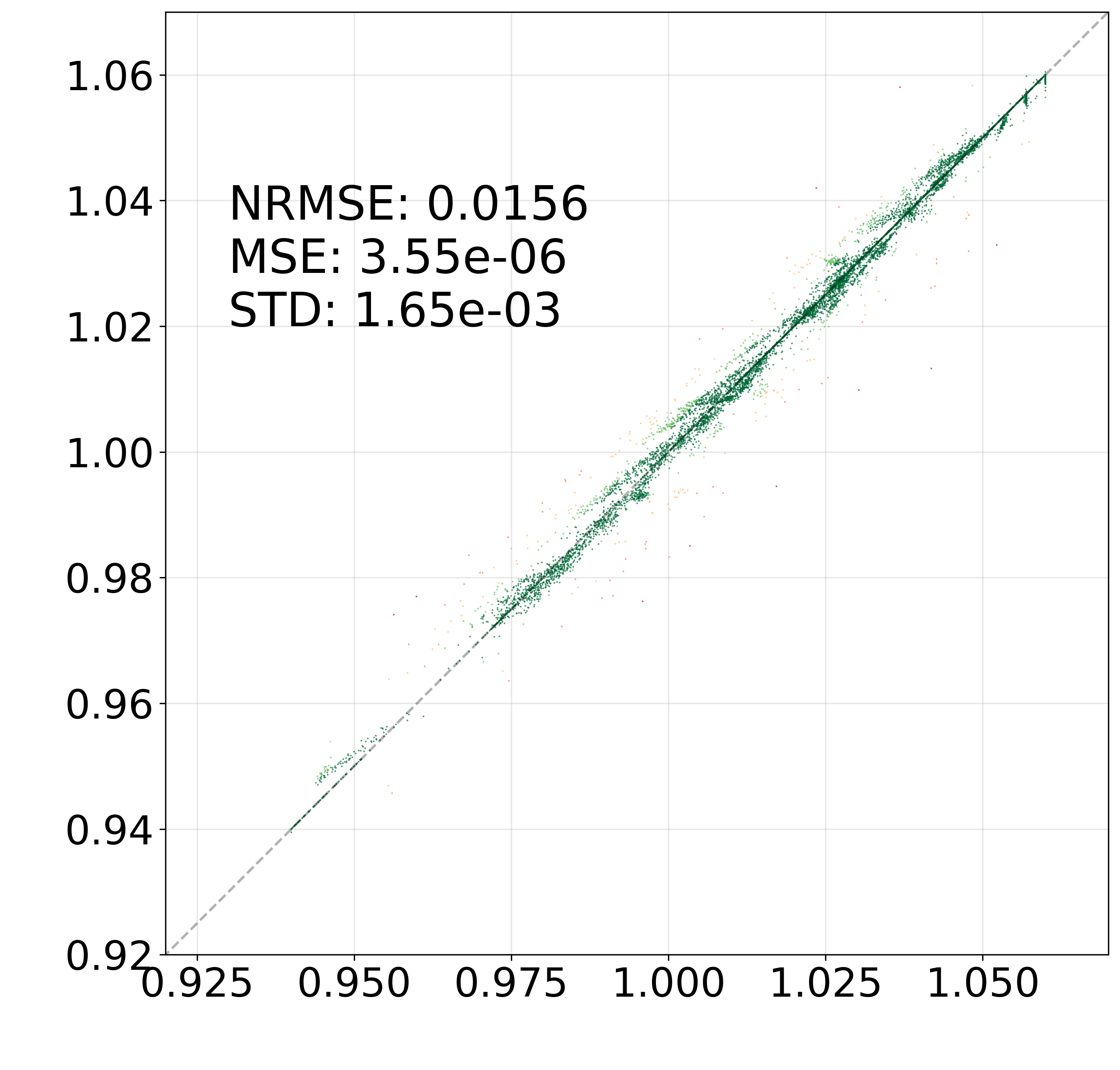}%
        \label{fig:sub4}%
    }%
    \caption{Voltage magnitude prediction vs. ground truth scatter plots for the IEEE 118-bus network, evaluated under four scenarios. The model was trained exclusively on the \textit{mid} scenario.}
    \label{fig:PF-results}
\end{figure*}

The ablation study demonstrates that not each architectural component contributes meaningfully to overall accuracy. Removal of global mean pooling produces the largest accuracy drop, increasing average NRMSE by approximately 16\% relative to the baseline (from 0.0031 to 0.0036), underscoring its importance for tasks such as OPF that require awareness of system-wide constraints. The local residual connection has negligible effect on in-distribution accuracy but slightly improves generalization to unseen scenarios, approximately 2\%. In contrast, removing the root residual $h^{(0)}$, the skip connection to the initial node encoding, reduces in-distribution accuracy, by 6.54\%, while improving out-of-distribution performance, 4,9\%. This confirms that residual connections are critical for mitigating the oversmoothing that degrades deep GNN representations, and that their benefit is most apparent when the model is evaluated on scenarios outside the training distribution. Contrary to expectation, removal of Layer Normalization marginally improves accuracy in both scenarios, despite its expected role in stabilizing training and preventing weight explosion. This suggests that, at the scale of the IEEE test systems used here, the regularization provided by Layer Normalization may be unnecessary, though it may become more important as the architecture is scaled to larger networks.

Increasing the hidden dimension consistently improves prediction accuracy up to a size of 128, beyond which gains diminish while inference time continues to grow. The number of GNN layers has a less consistent effect: additional layers improve in-distribution accuracy but degrade generalization to unseen scenarios, likely due to overfitting, while also increasing inference time. These trade-offs motivate the selection of 4 layers and a hidden size of 128 as the optimal configuration.

The relative inference time (test inference time/baseline inference time) reported in Table \ref{tab:comparison} scales moderately with model depth and hidden dimension, with the selected configuration (4 layers, hidden size 128) achieving a favorable accuracy–efficiency trade-off. 

The generalization results in Fig.~\ref{fig:PF-results} show that a model trained exclusively on the \textit{mid} scenario retains acceptable accuracy, with MSE ranging from 1.34e-7 to 3.55e-6,  when evaluated on the \textit{narrow}, \textit{wide}, and \textit{high-topo} datasets, including load levels up to 250\% of nominal and topologies modified by branch additions and removals, none of which were seen during training. This out-of-distribution robustness is a key property for operational deployment, where real-time grid conditions may deviate significantly from historical operating points. Notably, the unified model achieves accuracy comparable to, and in some cases surpassing, task-specific GNN solvers \cite{lin_powerflownet_2024,varbella_physics-informed_2024,arowolo_towards_2025}, despite being trained across three simultaneous tasks; however, direct comparison is complicated by differences in evaluation metrics (MSE, R², NRMSE, TRMAE) and benchmark networks across studies.

\section{Conclusion}
\label{sec:conclusion}

This paper has demonstrated that a single Heterogeneous ResGated GCN architecture can simultaneously solve \gls{pf}, \gls{opf}, and \gls{se} to accuracy comparable to task-specific models, while requiring only one training pipeline. By integrating line impedance parameters (R,X,B) directly into the message-passing mechanism and augmenting node updates with a global context MLP, the architecture captures both local \gls{kcl} physics and system-wide optimization constraints within a unified representation.

Our results demonstrate that this multi-task framework achieves high precision on PF and SE while producing feasible solutions within the constrained optimization setting of OPF. Specifically, the selected configuration achieves average NRMSE values below 0.004 (Table~\ref{tab:comparison}) across all predicted quantities on the \textit{mid} scenario, and generalizes to unseen loading conditions and topological configurations without retraining. The performance of the model is comparable to state-of-the-art task-specific GNNs, yet it offers superior versatility by eliminating the need for redundant, task-isolated architectures. Furthermore, the inclusion of topological perturbations and extreme load scaling (up to 250\%) supports robustness to grid reconfigurations and stressed operating conditions.

Future research is going to work to move the current implementation toward foundational models for power systems. This work establishes that a single GNN-based architecture can internalize shared physical principles, motivating future large-scale pre-training on global utility datasets. Such advancements will be critical for the autonomous operation and real-time control of the next generation of highly dynamic and decentralized power grids.

\subsection{Towards a Foundational Model for Power Systems}
The proposed GNN architecture demonstrates the capability to solve PF, OPF, and SE within a single unified model (Table \ref{tab:comparison}). We characterize this, however, not as a fully realized foundational model, but as an initial step which can internalize the shared physics of these three fundamental problems. This work advances three key prerequisites toward that goal:
\begin{enumerate}[leftmargin=*]
    \item \textit{Multi-Task Consolidation:} By merging descriptive tasks (PF, SE) with prescriptive optimization (OPF), we move away from specialized models toward a general-purpose grid solver.
    \item \textit{Topological Invariance via ResGated GCN:} The use of edge-conditioned message-passing layers, with strong graph-discriminative power, ensures that the model learns the underlying physics of \gls{kcl} rather than the specific geometry of a single test case, a prerequisite for foundational transfer learning.
    \item \textit{Physics-Informed Latent Space:} The integration of global context and deep residual paths allows the model to handle the high-dimensional complexity inherent in the global energy landscape.
\end{enumerate}
Consequently, this research serves as a proof of concept for a unified foundational power system model. Scaling this framework to encompass thousands of heterogeneous utility-scale systems would represent a concrete step toward a truly foundational model for autonomous power system operations.

\bibliography{biblio}          

\end{document}